\documentclass[fleqn,usenatbib]{mnras}

\usepackage{newtxtext,newtxmath}
\usepackage[T1]{fontenc}
\usepackage{ae,aecompl}
\usepackage{url}

\usepackage{graphicx}	% Including figure files
\usepackage{amsmath}	% Advanced maths commands
\usepackage{amssymb}	% Extra maths symbols
\usepackage{longtable}

\newcommand{\nup}{$\nu_{\rm peak}^S$}

\newcommand{\en}{E_{\nu}}
\title[AGN outflows as neutrino sources]{AGN outflows as neutrino sources: an observational test}

 \author[P. Padovani et al.]{P. Padovani$^{1,2}$\thanks{E-mail:
ppadovan@eso.org}, A. Turcati$^3$, E. Resconi$^3$\\
$^{1}$European Southern Observatory, Karl-Schwarzschild-Str. 2,
D-85748 Garching bei M\"unchen, Germany\\
$^{2}$Associated to INAF - Osservatorio Astronomico di Roma, via Frascati 33,
I-00040
Monteporzio Catone, Italy\\
$^{3}$Technische Universit{\"a}t M{\"u}nchen, Physik-Department, James-Frank-Str. 1, 
D-85748 Garching bei M{\"u}nchen, Germany\\
}

\date{Accepted XXX. Received YYY; in original form ZZZ}

\pubyear{2018}

\begin{document}
\label{firstpage}
\pagerange{\pageref{firstpage}--\pageref{lastpage}}
\maketitle

% Abstract of the paper
\begin{abstract}
We test the recently proposed idea that outflows associated with Active Galactic Nuclei (AGN) could 
be neutrino emitters in two complementary ways. First, we cross-correlate a list of 94 ``bona fide'' 
AGN outflows with the most complete and updated repository of IceCube
neutrinos currently publicly available, assembled by us for this purpose. It turns out that 
AGN with outflows matched to an IceCube neutrino have outflow and kinetic energy rates, and 
bolometric powers larger than those of AGN with outflows not matched to
neutrinos. Second, we carry out a statistical analysis on a catalogue of 
[O\,\textsc{iii}]~$\lambda5007$ line profiles using a sample of 23,264 AGN at $z < 0.4$, 
a sub-sample of which includes mostly possible outflows sources.  
We find no significant evidence of an association between the AGN and the IceCube events, 
although we get the smallest p-values ($\sim 6$ and 18 per cent respectively, pre-trial) for 
relatively high velocities and luminosities. Our results are consistent with a scenario 
where AGN outflows are neutrino emitters but at present do not provide a significant signal. This
can be tested with better statistics and source stacking. A predominant role of AGN outflows in explaining 
the IceCube data appears in any case to be ruled out. 
\end{abstract}

% Select between one and six entries from the list of approved keywords.
% Don't make up new ones.

%%%%%%%%%%%%%%%%%%%%%%%%%%%%%%%%%%%%%%%%%%%%%%%%%%

%%%%%%%%%%%%%%%%% BODY OF PAPER %%%%%%%%%%%%%%%%%%

\section{Introduction}\label{sec:Introduction}

The IceCube South Pole Neutrino
Observatory\footnote{\url{http://icecube.wisc.edu}} has reported in the
past few years the first observations of high-energy astrophysical
neutrinos\footnote{In this paper neutrino means both neutrino and
  antineutrino.}
\citep{2013PhRvL.111b1103A,ICECube13,ICECube14,ICECube15_1}. Recently, it
has confirmed and strengthened these observations by publishing a sample of
82 high-energy starting events (HESE) collected over six years and with a
deposited energy up to 2 PeV \citep{ICECube17_2}, which are inconsistent
with the hypothesis of purely terrestrial origin with very high
significance ($> 6.5\sigma$).

These HESE cover the whole sky and are mostly cascade-like events, which
can only be reconstructed with a spatial resolution in the tens of
degrees. There is also a complementary sample of through-going charged
current $\nu_\mu$ from the northern sky studied over a period of eight
years \citep{Aartsen2015,ICECube15_2,Aartsen2016,ICECube17_1}. These are
almost all track-like, meaning that their positions are known typically within 
one degree or less. 

Where are these neutrinos coming from? Since their sky distribution is
isotropic \cite[e.g.,][]{ICECube17_2} most of them need to have an
extragalactic origin, although a minor Galactic component cannot be
excluded. Many different scenarios for the astrophysical counterparts of
IceCube neutrinos have been put forward, including blazars, star-forming
galaxies, $\gamma$-ray bursts, galaxy clusters, and high-energy Galactic
sources \citep[see, e.g.][and references therein, for a comprehensive
  discussion]{Ahlers_2015}. Of these, blazars are so far the most
supported by the data.
  
Blazars are AGN \citep[see][for a recent 
  review]{Padovani_2017} having a jet at a small angle with respect to the
line of sight. The jet is highly relativistic and contains particles moving
in a magnetic field emitting non-thermal radiation over the whole
electromagnetic spectrum \citep{UP95,Padovani_2017}. Blazars come in two
main flavours: BL Lacertae objects (BL Lacs) and flat-spectrum radio
quasars (FSRQ). These differ mostly in their optical spectra, with the
latter displaying strong, broad, quasar-like emission lines and the former
instead having optical spectra with at most weak emission lines, sometimes
exhibiting absorption features, and in many cases totally featureless. The
possibility that blazars could be high-energy neutrino sources goes back to
long before the IceCube detections and has since been investigated in a
number of papers
\citep[e.g.,][]{mannheim95,halzen97,mueckeetal03,Pad_2014,Petro_2015,tav15}.

In particular, \cite{Padovani_2016} have correlated the second catalogue of
hard {\it Fermi}-LAT sources (2FHL) ($E>50\:\mathrm{GeV}$, 360 sources of
various types, mainly blazars, \citealt{2FHL}), and two other catalogues with
the publicly available high-energy neutrino sample detected by IceCube
(including only events with energy and median angular error $\ge 60$ TeV
and $\le 20^{\circ}$ respectively and covering the first four years of HESE
data).  The chance probability scanning over the $\gamma$-ray flux was 0.4
per cent, which becomes 1.4 per cent ($2.2\sigma$) by evaluating the impact
of the trials \citep[][see also Sect. \ref{sec:stat}]{Resconi_2017}.  This applies only to high-energy
peaked (HBL) blazars (sources with the peak of the synchrotron emission
\nup~$> 10^{15}$~Hz; \citealt{padgio95}) and appears to be strongly
dependent on $\gamma$-ray flux. The fraction of the IceCube signal
explained by HBL is however only $\sim 10 - 20$ per cent, which agrees with
the results of \cite{Pad_2015}, who have calculated the {\it cumulative}
neutrino emission from BL Lacs. Within the so-called {\it blazar simplified
  view} \citep[e.g.][]{paper1} and by adding a hadronic component from
\cite{Petro_2015} for neutrino production, BL Lacs as a class were in fact
shown to be able to explain the neutrino background seen by IceCube above
$\sim 0.5$ PeV while only contributing on average $\sim 10$ per cent at
lower energies. This is consistent with \cite{Aartsen2017}, who by
searching for cumulative neutrino emission from blazars in the second {\it
  Fermi}-LAT AGN (2LAC) catalogue, have constrained the maximum
contribution of 2LAC blazars to the observed astrophysical neutrino flux to
$< 27$ per cent. Similar results have been obtained by \cite{ICECube17_3}
using three more recent catalogues, a larger IceCube sample, and a range 
of $\gamma$-ray spectral shapes.

In what is so far the most significant result\footnote{\cite{Emig_2015}
  found a hint (p-value $\sim 0.3$ per cent) of an association between the
  37 IceCube neutrinos detected in the first three years of operation and a
  set of $\gamma$-ray detected starburst galaxies and star-forming regions
  in the Galactic neighbourhood. However, no trial correction was done (so
  the p-value is only a lower limit) and most of their sources appear to
  fail the ``energetic'' test suggested by \cite{Pad_2014}, i.e. their
  extrapolated $\gamma$-ray spectra fall well below the neutrino flux of
  the corresponding IceCube event.}  \cite{Resconi_2017} have presented a
strong hint of a connection between HBL, IceCube neutrinos, and ultra
high-energy cosmic rays (UHECRs; $E \ge 52 \times 10^{18}$ eV) with a
probability $\sim 0.18$ per cent ($2.9\sigma)$ after compensation for all
the considered trials. Even in this case, HBL can account only for $\approx
10$ per cent of the UHECR signal.

It is interesting to note that none of the possible neutrino counterparts
in \cite{Padovani_2016} and \cite{Resconi_2017} are tracks, as they are all
cascade-like. And indeed \cite{Palladino_2017} did not find a significant
correlation between 2FHL BL Lacs and 29 IceCube tracks. This indicates that
by using tracks we are still not sensitive to the HBL neutrino signal, as
also expected from the fact that tracks trace only about 1/6 of the
astrophysical signal under the assumption of a flavour ratio $\nu_{\rm
  e}:\nu_\mu:\nu_\tau = 1:1:1$ \citep[as pointed out
  by][]{Padovani_2016}. Very recently, however, \cite{Lucarelli_2017} have found
  a transient $\gamma$-ray ($> 100$ MeV) source positionally coincident with an 
  IceCube track with a post-trial significance $\sim 4\sigma$ and possibly associated 
  with an HBL. Moreover, $\gamma$-ray emission from
{\it Fermi}, AGILE, and MAGIC
\citep{2017ATel10791....1T,2017ATel10801....1L,2017ATel10817....1M} has
been detected from a BL Lac (with \nup~close to HBL values) inside the
error region of an IceCube track (\citealt{2017GCN.21916....1K}; see also
Padovani et al., in prep.). These could
be the {\it first} observed electromagnetic counterparts of IceCube
neutrinos.

In summary, while the evidence for HBL as neutrino emitters is getting
stronger, it is also clear that these sources cannot explain the whole
IceCube signal, which leaves room for other astrophysical components.

The purpose of this paper is to consider a new, possible class of neutrino
emitters: AGN outflows. Many AGN show evidence for large-scale outflows of
matter driven by the central black hole \citep[see, e.g.,][and references
  therein]{Harrison_2017,Fiore_2017}. These can reach semi-relativistic
speeds of up to $\sim$ 50,000 km s$^{-1}$ that can drive a shock that
accelerates and sweeps up matter \citep[e.g.,][]{King_2015}. The protons
accelerated by these shocks can generate $\gamma$-ray emission via
collisions with protons in the interstellar medium by producing neutral and
charged pions. The former decay into two $\gamma$-rays ($\pi^0 \rightarrow
\gamma + \gamma$) while the latter decay into secondary electrons,
positrons, and neutrinos. \cite{Wang_2016} and \cite{Lamastra_2017} have
shown that, using two different approaches and assuming that {\it all} AGN
have outflows, the neutrino emission from such outflows could explain the
{\it whole} IceCube signal. \cite{Liu_2018}, on the other hand, by including 
adiabatic losses not taken into account by other studies, have ruled out 
a dominant contribution to the IceCube flux from AGN outflows, which might 
contribute only at the $\approx 20$ per cent level. 
We stress that AGN outflows are not simply
(yet) another feature related to the central black hole but also play a
major role on galaxy scales through the so-called AGN feedback
\citep[e.g.,][for a review]{Fabian_2012}. This happens through the
interaction between the accretion-related radiation produced by the black
hole and gas in the host galaxy, which might sweep the galaxy bulge clean
of interstellar gas, put an end to star formation and, through lack of
fuel, also starve to death the AGN. Such a feedback mechanism explains in a
natural way the observed scaling relationship between the central black
hole and the host galaxy bulge mass.

The AGN outflow neutrino scenario has not been tested quantitatively so far. This is
what we plan to do in this paper by taking advantage of the larger neutrino
samples recently provided by the IceCube Collaboration (including both
cascades and tracks) and using two complementary approaches. Namely we want
to: (1) investigate the possible link between IceCube neutrinos and a very
comprehensive list of ``bona fide'' AGN displaying outflows; (2) study the
possible connection between IceCube neutrinos and a large catalogue of AGN
with optical spectral line information, which potentially includes
sources exhibiting outflows.

\onecolumn
\begin{longtable}{lrccrc}
\caption{IceCube events}\\
\hline
{IceCube ID} & {Dep. Energy} & {RA (2000)} & {Dec (2000)} & {Median Angular Error} & {Topology}\\ 
	& (TeV) & & & (deg) & \\
\hline
\endhead
\hline
\endfoot
HES3  & 75.7 	  & 08 31 36 & $-$31 12 00 & 1.4 & Track \\
HES4  & 159.1	  & 11 17 59 & $-$51 12 00 & 7.1 & Cascade \\
HES5  & 68.7 	  & 07 22 23 & $-$00 24 00 & 1.2 & Track \\
HES9  & 60.8 	  & 10 05 11 & +33 36 00 & 16.5 & Cascade \\
HES10 & 93.5 	  & 00 19 59 & $-$29 24 00 & 8.1 & Cascade \\
HES11 & 85.0	  & 10 21 12 & $-$08 54 00 & 16.7 & Cascade \\
HES12 & 100.2 	  & 19 44 24 & $-$52 48 00 & 9.8 & Cascade \\
HES13 & 243.1 	  & 04 31 36 & +40 18 00 & 1.2 & Track \\
HES14 & 1001.4   & 17 42 23 & $-$27 54 00 & 13.2 & Cascade \\
HES17 & 192.2 	  & 16 29 35 & +14 30 00 & 11.6 & Cascade \\
HES19 & 68.8 	  & 05 07 35 & $-$59 42 00 & 9.7 & Cascade \\
HES20 & 1097.7   & 02 33 11 & $-$67 12 00 & 10.7 & Cascade \\
HES22 & 211.2 	  & 19 34 48 & $-$22 06 00 & 12.1 & Cascade \\
HES23 & 79.1 	  & 13 54 47 & $-$13 12 00 & 1.9 & Track \\
HES26 & 202.1 	  & 09 33 35 & +22 42 00 & 11.8 & Cascade \\
HES30 & 123.8 	  & 06 52 47 & $-$82 42 00 & 8.0 & Cascade \\
HES33 & 370.2 	  & 19 30 00 & +07 48 00 & 13.5 & Cascade \\
HES35 & 1928.0   & 13 53 35 & $-$55 48 00 & 15.9 & Cascade \\
HES38 & 193.0 	  & 06 13 11 & +14 00 00 & 1.2 & Track \\
HES39 & 97.5 	  & 07 04 47 & $-$17 54 00 & 14.2 & Cascade \\
HES40 & 151.4 	  & 09 35 35 & $-$48 30 00 & 11.7 & Cascade \\
HES41 & 84.2 	  & 04 24 23 & +03 18 00 & 11.1 & Cascade \\
HES44 & 81.4 	  & 22 26 48 &	+00 00 00 & 1.2 & Track \\
HES45 & 413.6 	  & 14 35 59 & $-$86 18 00 & 1.2 & Track \\
HES46 & 152.0 	  & 10 01 59 & $-$22 18 00 & 7.6 & Cascade \\
HES47 & 71.5 	  & 13 57 35 & +67 24 00 & 1.2 & Track \\
HES48 & 100.8 	  & 14 11 59 & $-$33 12 00 & 8.1 & Cascade \\
HES51 & 63.7 	  & 05 54 23 & +54 00 00 & 6.5 & Cascade \\
HES52 & 152.2 	  & 16 51 11 & $-$54 00 00 & 7.8 & Cascade \\
HES56 & 104.2 	  & 18 41 59 & $-$50 06 00 & 6.5 & Cascade \\
HES57 & 132.1 	  & 08 11 59 & $-$42 12 00 & 14.4 & Cascade \\
HES59 & 124.6 	  & 04 13 11 & $-$03 54 00 & 8.8 & Cascade \\
HES60 & 93.0 	  & 02 10 48 & $-$37 54 00 & 13.3 & Cascade \\
HES62 & 75.8 	  & 12 31 35 & +13 18 00 & 1.3 & Track \\
HES63 & 97.4 	  & 10 39 59 & +06 30 00 & 1.2 & Track \\
HES64 & 70.8 	  & 09 37 59 & $-$27 18 00 & 10.6 & Cascade \\
HES66 & 84.2 	  & 08 34 47 & +38 18 00 & 18.3 & Cascade \\
HES67 & 165.7 	  & 22 22 47 & +03 00 00 & 7.0 & Cascade \\
HES70 & 98.8 	  & 06 15 35 & $-$33 30 00 & 12.3 & Cascade \\
HES71 & 73.5 	  & 05 22 47 & $-$20 48 00 & 1.2 & Track \\
HES74 & 71.3 	  & 22 44 00 & $-$00 54 00 & 12.7 & Cascade \\
HES75 & 164.0 	  & 17 15 59 & +70 30 00 & 13.1 & Cascade \\
HES76 & 126.3 	  & 16 00 47 & $-$00 24 00 & 1.2 & Track \\
HES79 & 158.2 	  & 01 38 23 & $-$11 06 00 & 14.6 & Cascade \\
HES80 & 85.6 	  & 09 46 24 & $-$03 36 00 & 16.1 & Cascade \\
HES81 & 151.8 	  & 03 00 00 & $-$79 24 00 & 13.5 & Cascade \\
HES82 & 159.3 	  & 16 03 35 & +09 24 00 & 1.2 & Track \\ \hline
AHES1 & 18883.62*  & 16 02 16 & +09 18 00 & 0.60$^{\dagger}$ & Track \\
AHES2 & 15814.74*  & 14 20 26 & $-$00 30 00 & 1.23$^{\dagger}$ & Track \\
AHES3 & 10431.02*  & 13 17 14 & $-$32 00 00 & 1.49$^{\dagger}$ & Track \\
AHES4 & 7546.05*   & 02 43 19 & +12 36 00 & 0.88$^{\dagger}$ & Track \\
AHES5 & 8858.64*   & 20 20 35 & $-$26 36 00 & 0.47$^{\dagger}$ & Track \\
AHES6 & 8685.07*   & 14 47 11 & $-$26 00 00 & 2.40$^{\dagger}$ & Track \\
AHES7 & 13906.14*  & 10 51 26 & $-$15 23 59 & 1.94$^{\dagger}$ & Track \\ \hline
DIF1  & 480    & 01 58 20 & +01 13 47 & 0.31$^{\dagger}$ & Track \\
DIF2  & 250    & 19 52 50 & +11 44 23 & 0.45$^{\dagger}$ & Track \\
DIF3  & 340    & 22 59 43 & +23 34 47 & 3.06$^{\dagger}$ & Track \\
DIF4  & 260    & 09 24 59 & +47 48 00 & 0.43$^{\dagger}$ & Track \\
DIF5  & 230    & 20 27 50 & +21 00 00 & 2.13$^{\dagger}$ & Track \\
DIF6  & 770    & 16 47 59 & +15 12 35 & 10.73$^{\dagger}$ & Track \\
DIF7  & 460    & 17 45 09 & +13 24 00 & 0.54$^{\dagger}$ & Track \\
DIF8  & 660    & 22 04 19 & +11 05 23 & 0.55$^{\dagger}$ & Track \\
DIF9  & 950    & 05 55 47 & +00 30 00 & 0.39$^{\dagger}$ & Track \\
DIF10 & 520    & 19 03 48 & +03 09 00 & 1.09$^{\dagger}$ & Track \\
DIF11 & 240    & 20 30 50 & +01 01 47 & 0.37$^{\dagger}$ & Track \\
DIF12 & 300    & 15 40 31 & +20 18 00 & 1.71$^{\dagger}$ & Track \\
DIF13 & 210    & 18 08 52 & +35 33 00 & 0.85$^{\dagger}$ & Track \\
DIF14 & 210    & 21 02 38 & +05 17 23 & 5.21$^{\dagger}$ & Track \\
DIF15 & 300    & 14 51 28 & +01 52 11 & 3.53$^{\dagger}$ & Track \\
DIF16 & 660    & 02 26 36 & +19 06 00 & 1.96$^{\dagger}$ & Track \\
DIF17 & 200    & 13 14 57 & +31 57 35 & 0.96$^{\dagger}$ & Track \\
DIF18 & 260    & 22 00 24 & +01 34 12 & 0.61$^{\dagger}$ & Track \\
DIF19 & 210    & 13 40 26 & $-$02 23 24 & 0.54$^{\dagger}$ & Track \\
DIF20 & 750    & 11 18 26 & +28 02 23 & 0.85$^{\dagger}$ & Track \\
DIF22 & 400    & 14 59 33 & $-$04 26 23 & 1.05$^{\dagger}$ & Track \\
DIF23 & 390    & 02 11 45 & +10 12 00 & 0.52$^{\dagger}$ & Track \\
DIF24 & 850    & 19 33 09 & +32 49 11 & 0.56$^{\dagger}$ & Track \\
DIF25 & 400    & 23 17 33 & +18 03 00 & 2.70$^{\dagger}$ & Track \\
DIF26 & 340    & 07 05 02 & +01 17 23 & 1.57$^{\dagger}$ & Track \\
DIF27 & 4450   & 07 22 31 & +11 25 12 & 0.37$^{\dagger}$ & Track \\
DIF28 & 210    & 06 41 55 & +04 33 35 & 1.08$^{\dagger}$ & Track \\
DIF29 & 240    & 06 06 23 & +12 10 48 & 0.40$^{\dagger}$ & Track \\
DIF30 & 300    & 21 41 59 & +26 06 00 & 1.62$^{\dagger}$ & Track \\
DIF31 & 380    & 21 53 35 & +06 00 00 & 0.55$^{\dagger}$ & Track \\
DIF32 & 220    & 08 55 59 & +28 00 00 & 0.45$^{\dagger}$ & Track \\
DIF33 & 230    & 13 10 23 & +19 53 59 & 2.33$^{\dagger}$ & Track \\
DIF34 & 740    & 05 05 11 & +12 36 00 & 0.66$^{\dagger}$ & Track \\
DIF35 & 380    & 01 02 23 & +15 36 00 & 0.53$^{\dagger}$ & Track \\
%DIF36 & 330.0    & 00 38 47 & +26 36 00 & ---  & Track \\
\hline
EHE1 & 15814.74* & 14 18 09 & $-$00 18 00 & 0.75 & Track \\
EHE2 & --- & 08 11 11 & $-$00 48 00 & 0.10 & Track \\
EHE3 & 100.00 & 03 06 19 & +15 00 00 & 0.78$^{\dagger}$ & Track \\
EHE4 & 120.00 & 06 33 11 & $-$15 00 00 & 1.18$^{\dagger}$ & Track \\
EHE5 & 120.00 & 05 09 43 & +05 41 59 & 0.83$^{\dagger}$ & Track \\
EHE6 & 230.00 & 22 39 59 & +07 24 00 & 0.47$^{\dagger}$ & Track \\
\hline
\multicolumn{6}{l}{\footnotesize * Deposited energy in photoelectronvolt units.}\\
\multicolumn{6}{l}{\footnotesize $^{\dagger}$ $90\%$ C.L. angular uncertainty}
\label{tab:ICE}
\end{longtable}
\twocolumn

Section 2 describes the neutrino and AGN outflow catalogues used in this
paper, while Section 3 gives our results, which are interpreted in Section 4. 
Section 5 summarises our conclusions. 
We use a $\Lambda$CDM cosmology with $H_0 = 70$ km
s$^{-1}$ Mpc$^{-1}$, $\Omega_m = 0.3$, and $\Omega_\Lambda = 0.7$. 
%\citep{kom11}.

\section{The catalogues}

\subsection{Neutrino lists}\label{sec:neutrino_list}

This work is based on the IceCube HESE published by
\cite{ICECube14,ICECube15_1,ICECube17_2}, which cover the first six years
of data (HES in Tab. \ref{tab:ICE}) and the $\nu_{\mu}$ selected from a
large sample of high-energy through-going muons by applying a 200 TeV
deposited energy threshold (DIF)
\citep[see][]{Aartsen2015,ICECube15_2,Aartsen2016,ICECube17_1}.  We also
include the neutrinos provided by the Astrophysical Multimessenger
Observatory Network (AMON), which include starting (AHES) and extremely
high-energy (EHE) events available on-line\footnote{See the lists at
  \url{https://gcn.gsfc.nasa.gov/amon_hese_events.html} and
  \url{https://gcn.gsfc.nasa.gov/amon_ehe_events.html}}. Given that these
lists are only partially up to date, for these neutrinos we gathered the
most recent information by checking the Gamma-ray Coordinates Network (GCN)
archive. We have also excluded two events: one which has been retracted and
another one, which was missing the angular error.
        
Following \cite{Pad_2014} we made the following two cuts to the HESE list:
(1) neutrino energy $\en \ge 60$ TeV, to reduce the residual atmospheric
background contamination, which might still be produced by mouns and
atmospheric neutrinos and concentrates in the low-energy part of spectrum
\citep[see Fig. 2 in][]{ICECube14}; (2) median angular error $\le
20^{\circ}$, to somewhat limit the number of possible counterparts. The
final list includes 47 HESE, 34 through-going $\nu_{\mu}$, 7 AHES, and 6
EHE, for a total of 94 IceCube events. These are listed in
Tab. \ref{tab:ICE}, which gives the ID, the deposited energy of the
neutrino, the coordinates, the median angular error or 90\% uncertainty in
degrees, and the event topology.

We stress that, to the best of our knowledge, Tab. \ref{tab:ICE} is the
{\it only} complete (modulo the two cuts made to the HESE list) and updated
repository of IceCube neutrinos currently publicly available.

\subsection{AGN outflow catalogues}

\subsubsection{The AGN outflow list}\label{sec:Fiore_cat}

\cite{Fiore_2017} have studied scaling relations between AGN properties,
host galaxy properties, and AGN outflows. To do so, they have assembled
from the literature observations of 94 distinct AGN with reliable massive
outflow detections, for which there was an estimate (or a robust limit) on
the physical size of the high velocity gas in the wind. As stressed by the
authors their sample is not complete and suffers from strong selection
biases, different for the various types of outflows. In particular, most
molecular winds and ultrafast outflows (UFOs) can only be studied locally
(typically at $z \lesssim 0.2$), ionised winds are found both at
low-redshift and at $z \sim 2-3$, while broad absorption line (BAL) sources
are at $z \sim 2-3$. Their list (see their Tab. B1) is therefore a very
comprehensive compilation of AGN outflows but does not represent a
well-defined sample with which to do statistical studies. As such, it is
therefore fully complementary to the SDSS catalogue, discussed below.

\subsubsection{The SDSS catalogue}

\cite{Mullaney_2013} have presented a catalogue of [O\,\textsc{iii}]~$\lambda5007$ 
line profiles using a sample of 23,264 AGN at $z < 0.4$ selected from the Sloan 
Digital Sky Survey (SDSS) DR7 
data base. These can be used to determine the kinematics of the 
kpc-scale emitting gas. Using their data we have computed the 
[O\,\textsc{iii}]~$\lambda5007$ flux-weighted average full width 
half-maximum (FWHM):

\begin{equation}
{\rm FWHM_{\rm Avg} = [(FWHM_{\rm broad}~F_{\rm broad})^2 + (FWHM_{\rm narr}~F_{\rm narr})^2]^{1/2}}
\end{equation} 

where ${\rm F_{\rm broad}}$ and ${\rm F_{\rm narr}}$ are the fractional
fluxes contained within the two fitted Gaussian components of the
[O\,\textsc{iii}]~$\lambda5007$ line, a broad and a narrow one. As
discussed by \cite{Mullaney_2013} it is better to use the average FWHM 
rather than a single component
(e.g., ${\rm FWHM_{\rm broad}}$) as this also allows inclusion of sources
for which the line is satisfactorily fitted by a single Gaussian. We note
that ${\rm FWHM_{\rm Avg}} > 500$ km s$^{-1}$ is the typical lower limit
generally adopted when selecting targets for follow-up (i.e., with integral
field unit [IFU]) spectroscopic studies of AGN outflows
\citep[e.g.,][]{Harrison_2014}. The \cite{Mullaney_2013} catalogue 
includes 17 per cent of sources above this value and is therefore a very good,
well-defined catalogue of AGN with possible outflows to be used for
statistical studies. \cite{Harrison_2014}, in fact, have presented IFU observations 
of 16 AGN selected from this sample (at the high ${\rm FWHM_{\rm Avg}}$ and 
$L_{[\ion{O}{iii}]}$ end) and have detected high-velocity outflows on kpc scales in all of them.  
Power is also another good outflow indicator. For this purpose we use below the {\it observed}
$L_{[\ion{O}{iii}]}$ in \cite{Mullaney_2013} (and not the {\it de-reddened} one as this
reaches some very large and unphysical values due to the V-band magnitude
extinction used).
  
\section{Results}

\subsection{AGN outflow list}

\begin{figure*}
%\vspace{0.8cm}
\includegraphics[height=7.0cm]{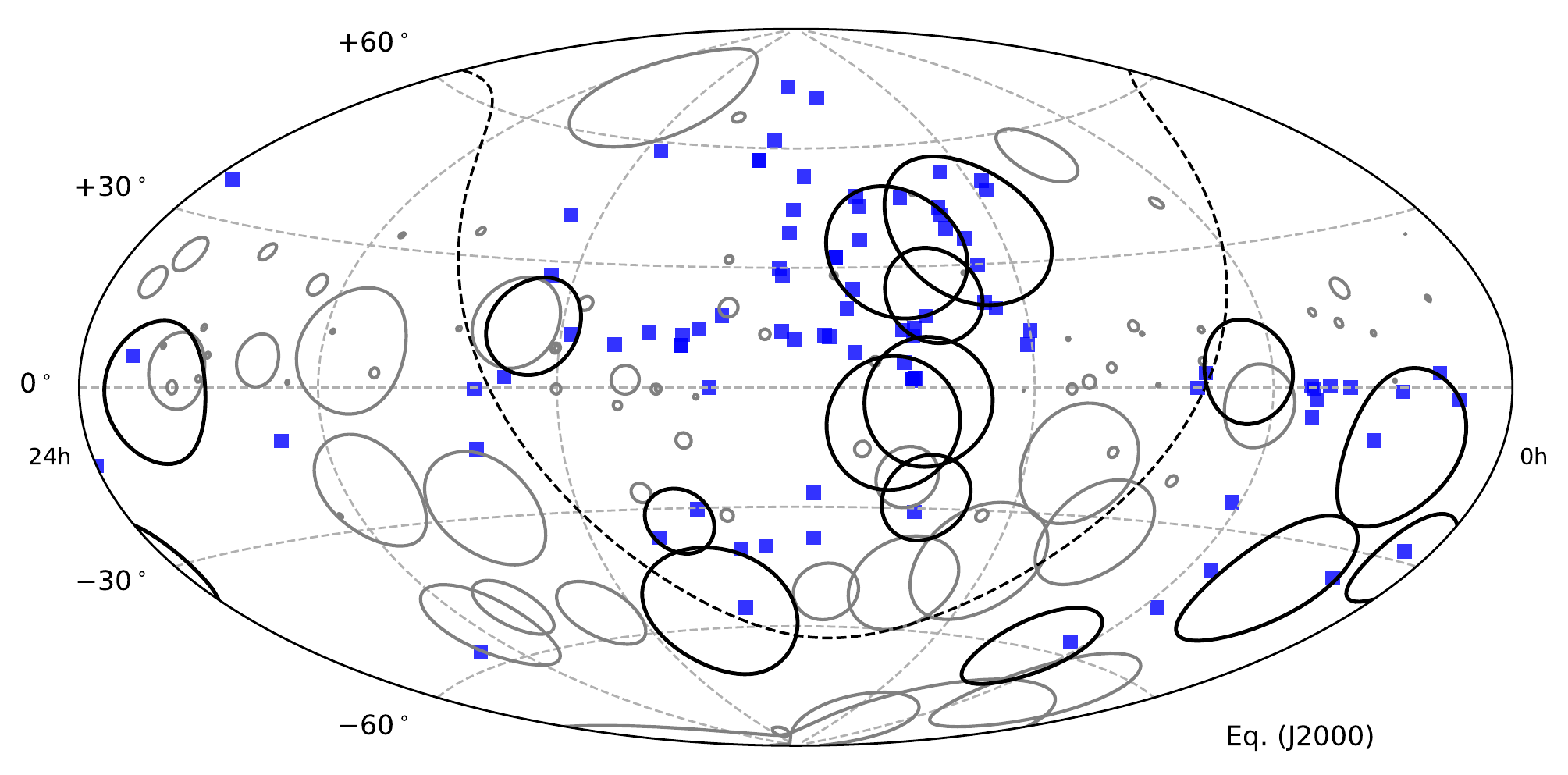}
\caption{Sky map in equatorial coordinates. Blue squares indicate the AGN
  with outflows from \protect\cite{Fiore_2017}, while IceCubes events are represented as circles with
  radius equal to the angular uncertainty. Black circles indicate neutrinos
  with a counterpart, grey circles neutrinos without counterparts. The
  dashed line represents the Galactic plane.}
\label{fig:Outflow_skymap}
\end{figure*}

We cross-correlated our neutrino list with the AGN outflow list of \cite{Fiore_2017}. 
An outflow counterpart was found within the given angular error for 
15/96 neutrino events, all HESE of the cascade type. These correspond to
45 entries, 9 of which were matched to multiple IceCube events for a total
of 36 distinct AGN. Our results are shown in
Tab. \ref{tab:AGN_counterparts}, which gives the IceCube ID, the AGN
counterpart's name and coordinates, the offset between the reconstructed
position of the IceCube event and the AGN one, the source redshift, the
outflow type, and the HBL listed as ``most probable'' matches in
\cite{Padovani_2016}. These are sources, which not only are within the
median error radius of an IceCube neutrino but for which a simple
extrapolation of their $\gamma$-ray spectral energy distribution (SED)
connects to the neutrino flux.

Fig. \ref{fig:Outflow_skymap} shows the positions of the AGN with outflow from 
\cite{Fiore_2017} (blue
squares) in equatorial coordinates, while IceCubes events are represented
as circles with radius equal to the angular error (with black colour
indicating neutrinos with a counterpart and grey neutrinos without). 
Given the statistical limitations of the AGN outflow list, discussed in Sect. 
\ref{sec:Fiore_cat}, we cannot test the statistical significance of the
neutrino - outflow matches.

\cite{Fiore_2017} have presented the basic properties of the AGN with
outflows discussed in their paper, deriving also in an homogenous way
physical quantities such as the mass outflow rate, $\dot{\rm M}_{\rm OF}$,
and the kinetic energy rate, $\dot{\rm E}_{\rm kin}$. These are the
instantaneous outflow rate of material at the edge of the outflow region
(see their eq. B.1 and B.2) and its kinetic power (equal to \textonehalf $~\dot{\rm
  M}_{\rm OF}~v{^2}_{\rm max}$, where $v_{\rm max}$ is the outflow maximum
velocity).  We have looked for possible parameter differences between
outflows with and without an IceCube match\footnote{The \cite{Fiore_2017}
  list includes multiple entries for some AGN, as outflows were detected in
  more than one way. We used a list of distinct sources by keeping the
  entry with the maximum value of $\dot{\rm M}_{\rm OF}$.}, finding three:
$\dot{\rm E}_{\rm kin}$, $\dot{\rm M}_{\rm OF}$, and bolometric power.

\begin{figure}
\includegraphics[height=8.7cm]{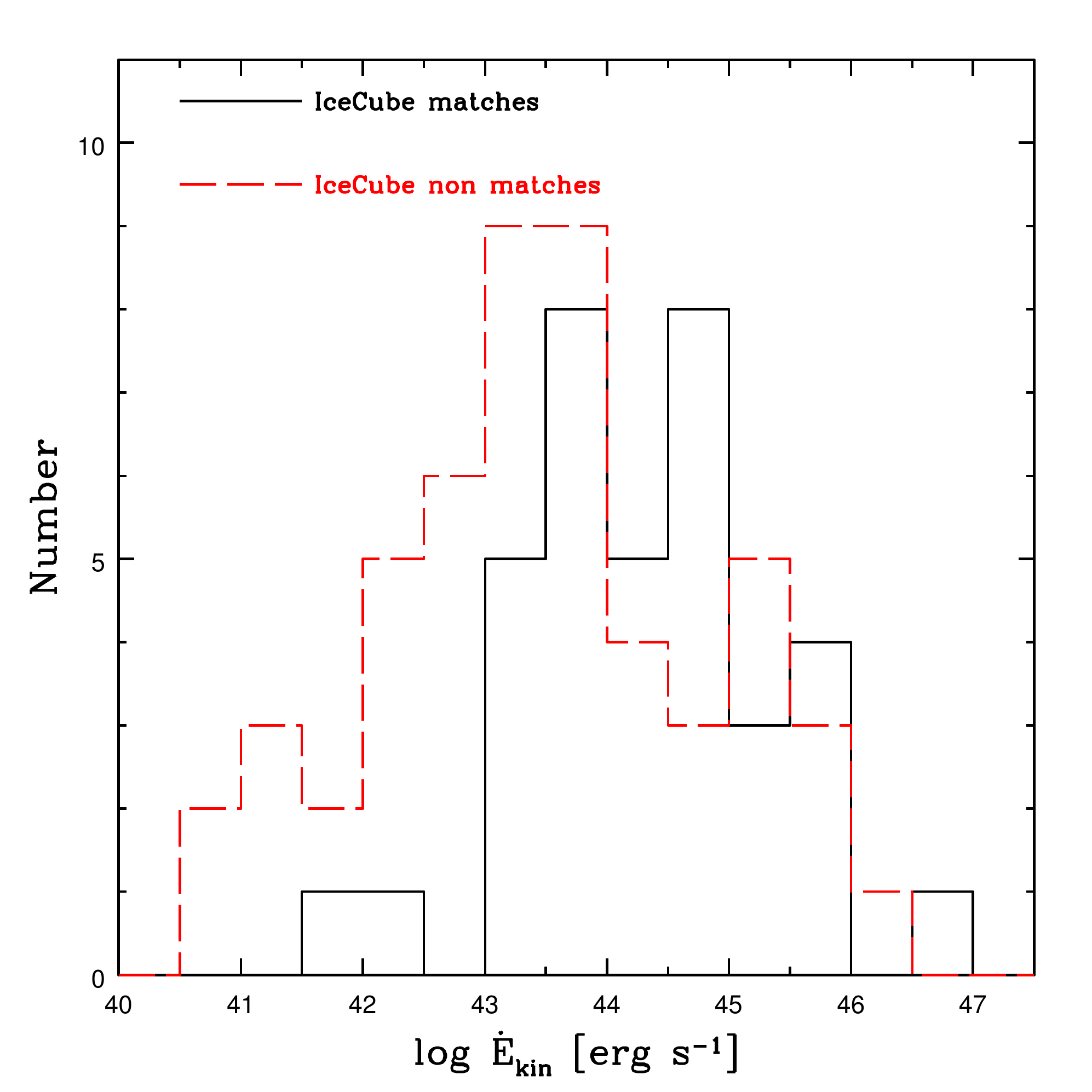}
%\includegraphics[height=8.7cm]{hist_ekin.eps}
%\vspace{0.8cm}
\caption{The distribution of the kinetic power for AGN outflows in the list
  of \protect\cite{Fiore_2017} with (solid black line) and without (red
  dashed line) IceCube counterparts.}
\label{fig:ekin}
\end{figure}

Figure \ref{fig:ekin} shows the distribution of $\dot{\rm E}_{\rm kin}$ for
AGN outflows with (solid black line) and without (red dashed line) IceCube
counterparts. The two distributions are significantly different ($P \sim
99.6$, $\sim 99.7$, and $>99.9$ per cent according to a Kolmogorov-Smirnov
[KS], Mann-Whitney-Wilcoxon [MWW], and Cramer test
respectively\footnote{Although the list is not complete and biased, none of
  the biases are neutrino related and therefore these tests are
  meaningful.}), with the outflows with IceCube matches having $\langle
\dot{\rm E}_{\rm kin} \rangle$ {\it larger} by a factor $\sim 7$ than the
corresponding value for outflows without IceCube matches.

\begin{figure}
\includegraphics[height=8.7cm]{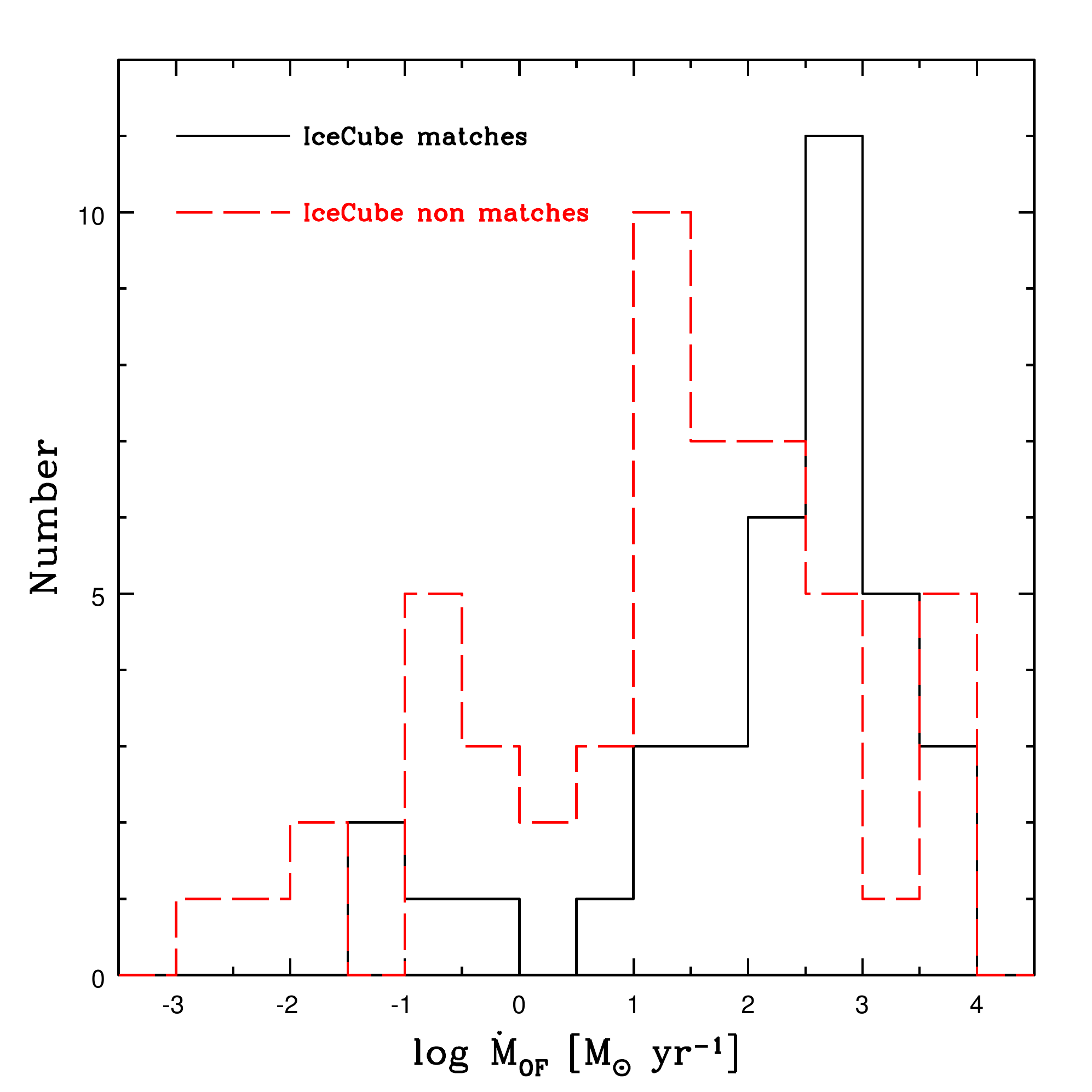}
%\includegraphics[height=8.7cm]{hist_mdot.eps}
%\vspace{0.8cm}
\caption{The distribution of the mass outflow rate for AGN outflows in the
  list of \protect\cite{Fiore_2017} with (solid black line) and without
  (red dashed line) IceCube counterparts.}
\label{fig:mdot}
\end{figure}

Figure \ref{fig:mdot} shows the distribution of $\dot{\rm M}_{\rm OF}$ for
AGN outflows with (solid black line) and without (red dashed line) IceCube
counterparts. The two distributions are significantly different ($P \sim
99.6$, $\sim 99.6$, and $\sim 99.8$ per cent according to a KS, MWW, and
Cramer test respectively), with the outflows with IceCube matches having
$\langle \dot{\rm M}_{\rm OF} \rangle$ {\it larger} by a factor $\sim 7$
than the corresponding value for outflows without IceCube matches.

Finally, the AGN bolometric power distributions for AGN outflows with and without IceCube
counterparts are marginally different ($P \sim 96.2$, $\sim 97.6$, and
$\sim 97.3$ per cent according to a KS, MWW, and Cramer test respectively),
with the outflows with IceCube matches having bolometric powers {\it
  larger} by a factor $\sim 4$ than the corresponding value for outflows
without IceCube matches.

As shown in Tab. \ref{tab:AGN_counterparts} four IceCube events have ``most probable'' 
matches with HBL, so one could argue that these events should be deleted from this list
as it is unlikely they might be associated with AGN outflows. 
If one does so all differences between AGN outflows with and without IceCube
counterparts disappear ($P < 89\%$). In practice, however, this is independent of the 
choice of the removed IceCube events but it is simply due to the fact that 11 AGN with 
relatively high values of $\dot{\rm E}_{\rm kin}$, $\dot{\rm M}_{\rm OF}$, and 
bolometric power are moved from one list to the other. 

Given this intriguing result, we have followed this up by using a 
well-defined and complete catalogue, that is the SDSS catalogue, 
to which we can apply a statistical test to study possible correlations. 
 
\subsection{SDSS catalogue}

\subsubsection{The statistical analysis}\label{sec:stat}

\begin{figure*}
\includegraphics[height=7.0cm]{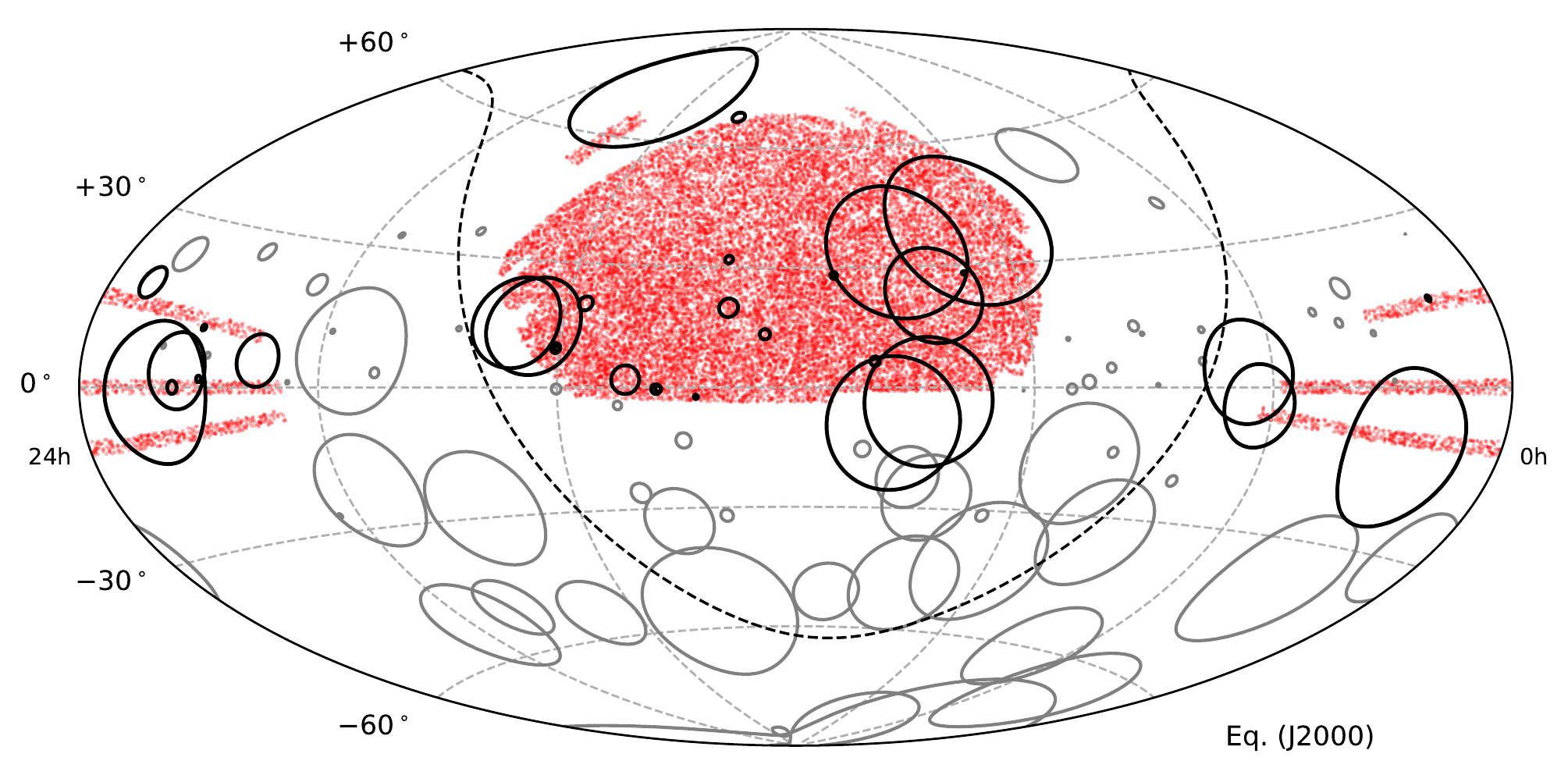}
%\includegraphics[height=7.0cm]{Alpaka_skymap.pdf}
%\vspace{0.8cm}
\caption{Sky map in equatorial coordinates. Red dots indicate the SDSS
  objects while IceCubes events are represented as circles with radius
  equal to the angular error. Black circles indicate neutrinos with a
  counterpart, grey circles neutrinos without counterparts. The dashed line
  represents the Galactic plane.}
\label{fig:SDSS_skymap}
\end{figure*}

\begin{figure*}
\includegraphics[height=8.0cm]{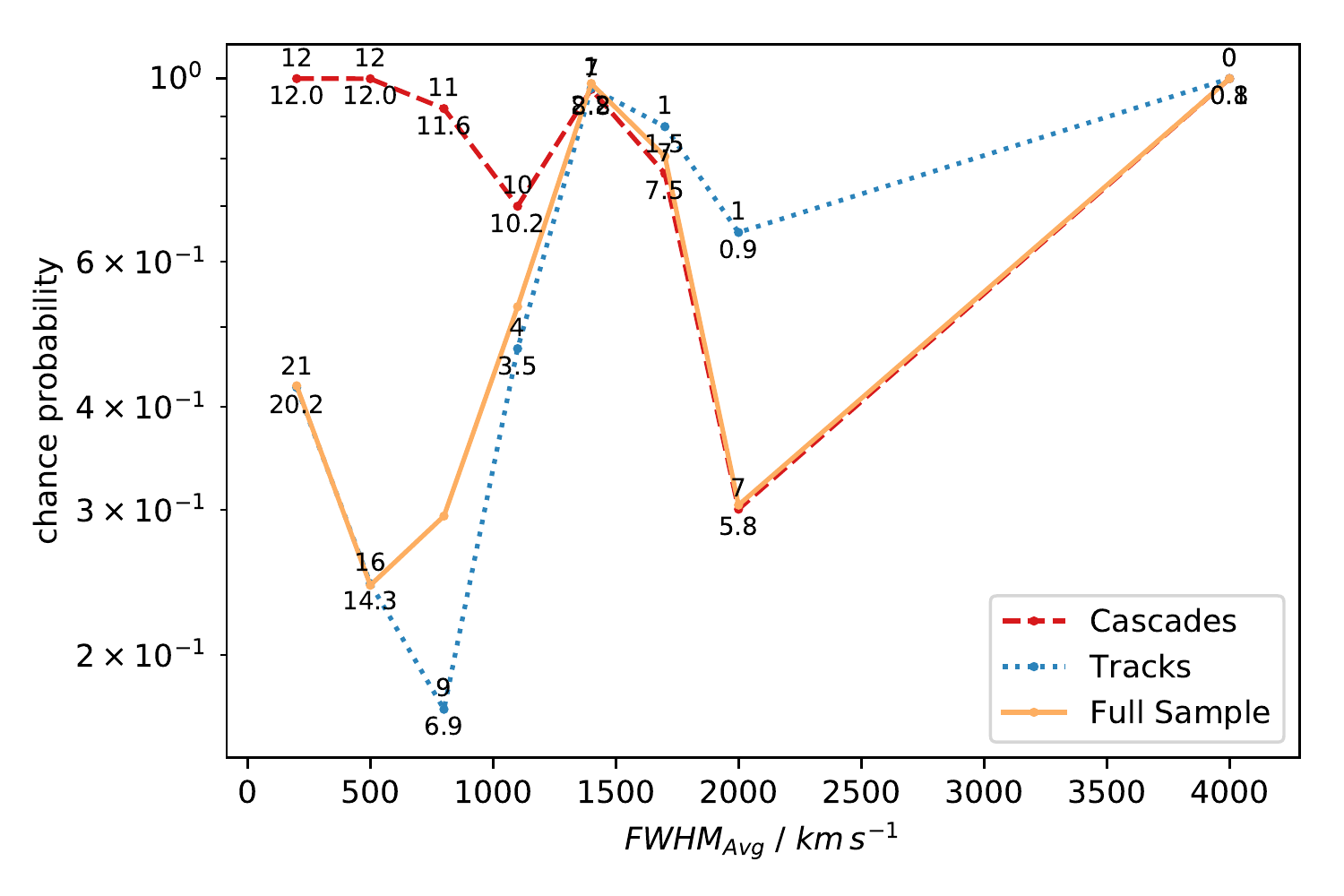}
%\includegraphics[height=8.0cm]{ALPAKA_FWHM_Final.pdf}
%\vspace{0.8cm}
\caption{The chance probability of association of SDSS AGN for objects
  having ${\rm FWHM_{\rm Avg}}$ larger than the value on the $x$-axis with
  all IceCube events (solid orange line), cascades (dashed red line), and
  tracks (dotted blue line). The numbers give the observed (above the
  points) and average random values (below the points) of $N_{{\nu}}$.
  A p-value $\sim 17$ per cent is reached for ${\rm FWHM_{\rm Avg}}
  \gtrsim 800$ km s$^{-1}$ for tracks.}
\label{fig:SDSS_FWHM}
\end{figure*}

\begin{table*}
%\begin{minipage}{110mm}
\caption{AGN with outflows within one median angular error radius from the positions of the ICeCube neutrinos.}
%\label{mathmode}
\begin{tabular}{@{}llllrlll}
ID &  Name &  RA (2000) & DEC (2000) &  offset & $z$ & outflow type  & Other ``most probable'' matches$^a$ \\
          &                    &   & &  (deg) & & &  \\
\hline  
 HES9         &       I08572$+$3915     &  09  00 25.4 &  $+$39  03 54 & 14.1& 0.05835   &  molecular \& ionised & MKN 421, 1ES 1011$+$496\\   
 HES9         &       I10565$+$2448     & 10 59 18.1 &  $+$24 32 34 & 14.9& 0.04311   &   molecular \& ionised &    ~~~~~~~~~~~    "    ~~~~~~~~~~~          "\\   
 HES9         &       I11119$+$3257     & 11 14 38.9 &  $+$32 41 33 & 14.5& 0.189     & molecular \& UFO &     ~~~~~~~~~~~    "    ~~~~~~~~~~~          "\\
 HES9         &       SDSSJ0900       &  09  00 33.5 &  $+$42 15 47 & 15.4& 3.297     &   ionised &     ~~~~~~~~~~~    "    ~~~~~~~~~~~          "\\
 HES9         &       SMMJ0943        &  09 43  04.1 &  $+$47  00 16 & 14.0& 3.351     &  ionised &     ~~~~~~~~~~~    "    ~~~~~~~~~~~          "\\
 HES9         &       SDSSJ1039       & 10 39 27.2 &  $+$45 12 15 & 13.3& 0.579     &  ionised &     ~~~~~~~~~~~    "    ~~~~~~~~~~~          "\\
 HES9         &       SDSSJ1040       & 10 40 14.4 &  $+$47 45 55 & 15.6& 0.486     &  ionised &     ~~~~~~~~~~~    "    ~~~~~~~~~~~          "\\
 HES9         &       QSO1044         & 10 44 59.6 &  $+$36 56  05 &  8.8& 0.7       &  BAL &     ~~~~~~~~~~~    "    ~~~~~~~~~~~          "\\
 HES10        &       2QZJ0028        &  00 28 30.4 & $-$28 17  06 &  2.2& 2.401   &   ionised &   H 2356$-$309        \\
 HES11        &       SDSSJ10100      & 10 10 43.4 &  $+$06 12  01 & 15.3& 0.0984    &    ionised &  \\   
 HES11        &       COS11363        & 10  00 28.7 &  $+$02 17 45 & 12.3& 2.1       &   ionised &  \\   
 HES11        &       XID2028         & 10  02 11.3 &   $+$01 37 07 & 11.5& 1.593     &    ionised & \\   
 HES11        &       XID5321         & 10  03  08.8 &  $+$02  09  04 & 11.9& 1.47      &   ionised & \\   
 HES11        &       XID5395         & 10  02 58.4 &  $+$02 10 14 & 12.0& 1.472     & ionised &   \\   
 HES11        &       MIRO20581       & 10  00  00.6 &  $+$02 15 31 & 12.3& 2.45      &   ionised &  \\   
 HES17        &       SDSSJ1549       & 15 49 38.7 &  $+$12 45  09 &  9.9& 2.367     &  ionised &  PG 1553$+$113\\  
 HES19        &       1H0419$-$577    &  04 26  00.7 & $-$57 12 01 &  6.0& 0.104   &  UFO &    \\   
 HES26        &       SDSSJ0945       &  09 45 21.3 &  $+$17 37 53 &  5.8& 0.1283    &   ionised &  \\   
 HES26        &       SDSSJ0958       &  09 58 16.9 &  $+$14 39 24 &  9.9& 0.1092    &    ionised & \\   
 HES35        &       I13120$-$5453   & 13 15  06.3 & $-$55  09 23 &  5.5& 0.03076 &  molecular &  \\   
 HES41        &       HB8905          &  05  07 36.4 &  $+$03  07 52 & 10.8&   2.48 &   ionised & 1ES 0414$+$009\\  
 HES48        &       I14378$-$3651   & 14 40 59.0 & $-$37  04 32 &  7.1& 0.06764    &  molecular &  \\   
 HES48        &       IC4329A         & 13 49 19.2 & $-$30 18 34 &  5.6&   0.016054 &  UFO &  \\   
 HES60        &       HE0109          &  01 11 43.5 & $-$35  03  01 & 12.2& 2.407    &    ionised &   \\   
 HES64        & MCG$-$5$-$23$-$16     &  09 47 40.1 & $-$30 56 55 &  4.2& 0.008486   &  UFO &   \\   
 HES66        &       I08572$+$3915$^b$     &  09 00 25.4 &  $+$39  03 54 &  5.1&  0.05835   &  molecular \& ionised & \\   
 HES66        &       SDSSJ0745       &  07 45 21.8 &  $+$47 34 36 & 12.9&   3.22   &   ionised &  \\   
 HES66        &       SDSSJ0900$^b$      &  09  00 33.5 &  $+$42 15 47 &  6.3&   3.297     &   ionised &   \\ 
 HES66        &       SMMJ0943$^b$        &  09 43  04.1 &  $+$47  00 16 & 15.2&  3.351     &  ionised &   \\   
 HES66        &       SDSSJ0841       &  08 41 30.8 &  $+$20 42 20 & 17.7&   0.641   &   ionised & \\   
 HES66        &       SDSSJ0842       &  08 42 34.9 &  $+$36 25  03 &  2.4&  0.561    &    ionised & \\   
 HES66        &       SDSSJ0858       &  08 58 29.6 &  $+$44 17 35 &  7.5&  0.454 & ionised &     \\   
 HES66        &       SDSSJ0838       &  08 38 17.0 &  $+$29 55 27 &  8.4&  2.043  &  BAL &  \\   
 HES66        &       Mrk79           &  07 42 32.8 &  $+$49 48 35 & 14.8&  0.022189   &  UFO &  \\   
 HES66        &       APM08279        &  08 31 41.7 &  $+$52 45 18 & 14.5&  3.91   &  UFO &  \\   
 HES74        &       I23060$+$0505     & 23  08 33.9 &   $+$05 21 30 &  8.8& 0.173   &  molecular &    \\   
 HES79        &       SDSSJ0149       &  01 49 32.5 &   $+$00 48  04 & 10.7& 0.567   &  ionised &  \\   
 HES79        &       SDSSJ0210       &  02 10 47.0 & $-$10  01 53 &  8.0&  0.54 &   ionised &  \\   
 HES79        &       Mrk279          &  00 52  08.9 &  $-$02 13  06 & 14.5&  0.030451  &  warm absorber  \\   
 HES80        &       SDSSJ10100$^c$      & 10 10 43.4 &   $+$06 12  01 & 11.5&    0.0984    &    ionised &  \\   
 HES80        &       COS11363$^c$        & 10  00 28.7 &   $+$02 17 45 &  6.9&    2.1       &   ionised &  \\   
 HES80        &       XID2028$^c$         & 10  02 11.3 &   $+$01 37  07 &  6.5&   1.593     &    ionised & \\   
 HES80        &       XID5321$^c$         & 10  03  08.8 &   $+$02  09  04 &  7.1&   1.47      &   ionised & \\   
 HES80        &       XID5395$^c$         & 10  02 58.4 &   $+$02 10 14 &  7.1&   1.472     & ionised &   \\  
 HES80        &       MIRO20581$^c$       & 10  00  00.6 &   $+$02 15 31 &  6.8&   2.45      &   ionised &  \\   
\hline
\multicolumn{8}{l}{\footnotesize $^a$ \cite{Padovani_2016}}\\
\multicolumn{8}{l}{\footnotesize $^b$ also counterpart of ID 9}\\
\multicolumn{8}{l}{\footnotesize $^c$ also counterpart of ID 11}\\
\label{tab:AGN_counterparts}
\end{tabular}
\end{table*}

To study the possible connection between the IceCube neutrinos and the
SDSS catalogue we follow the method of \cite{Padovani_2016}, which we
briefly summarise here.

We use the observable $N_{{\nu}}$ defined as the number of neutrino events
with at least one outflow counterpart found within the individual angular
uncertainty. We do not only consider the whole catalogue but we
additionally scan versus ${\rm FWHM_{\rm Avg}}$, $N_{{\nu}}({\rm FWHM_{\rm
    Avg}})$, and versus $[\ion{O}{iii}]$ luminosity,
$N_{{\nu}}(L_{[\ion{O}{iii}]})$. If only sources with higher outflow
velocities or luminosity are associated with IceCube events, such scans
will reveal a deviation from the randomised cases.

The chance probability $P_i(N_{{\nu}}({\rm FWHM_{\rm Avg}},i))$, or
equivalently $P_i(N_{{\nu}}(L_{[\ion{O}{iii}]},i)$), to observe a given
$N_{{\nu}}$ for sources with ${\rm FWHM_{\rm Avg}} \geq {\rm FWHM_{\rm
    Avg}},i$ is determined on an ensemble of typically $10^5$ randomised
maps. As discussed by \cite{Padovani_2016} scrambling on the neutrinos
right ascension does not conserve the total area sampled by the IceCube
error circles, resulting in a biased statistics. To correctly compare the
results of a random skymap with real data, in fact, the overlapping area
identified by the neutrino angular uncertainty and the portion of the sky
covered by the survey must be conserved in each random realisation.
This can be achieved by randomising the SDSS coordinates inside the portion
of the sky covered by the survey. This area has been approximated using an
HEALPix sky pixelisation with a total of $49152$ pixels, each covering
$0.84$ square degrees.

A p-value is then calculated for each of the bins ${\rm FWHM_{\rm Avg}},i$
($L_{[\ion{O}{iii}]},i$), for a total of 8 (11) p-values. When only
reporting the lowest p-value observed as a result of the analysis, a trial
correction for the ``{\it Look Elsewhere Effect}'' is needed 
\citep[e.g.,][]{2016ChPhC..40j0001P}
This stems from a simple fact: in the ideal case of 20 completely
independent tests, for example, one will observe one result more
significant than $\sim 2\sigma$ simply by chance. The final p-value can in this
case be trial corrected by multiplying it by the number of degrees of
freedom, i.e. the number of tests.  In the cases presented in this paper,
however, the bins with a lower value of our scanning parameter include also
the bins with a higher value, making our tests not independent.  The
analytical approximation for the trial correction is then no longer valid,
and one needs to study the distribution of the lowest p-value obtained from
the randomised cases. A trial corrected p-value can then be calculated as
the ratio between the number of randomised trials that produce a best
p-value at least as significant as the one given by the data, and the total
number of randomised trials.

Fig. \ref{fig:SDSS_skymap} shows the positions of the SDSS sources (red
dots) in equatorial coordinates, while IceCubes events are represented as
circles with radius equal to the angular error (with black colour
indicating neutrinos with a counterpart and grey neutrinos without).  Given
that we perform the randomization on the SDSS positions and due to the
large density of SDSS sources, it is apparent that IceCube events with
large angular errors (mostly cascades) will almost always give a match (even
when the sample gets smaller due to cuts on ${\rm FWHM_{\rm Avg}}$ or
$L_{[\ion{O}{iii}]}$: see below) and therefore by default cannot give a
signal. We therefore split the sample into cascades and tracks.

\subsubsection{Association probabilities}

Figure \ref{fig:SDSS_FWHM} shows the chance probability of association of
the SDSS sources with IceCube events for objects having ${\rm FWHM_{\rm
    Avg}}$ larger than the value on the $x$-axis. The dashed red line
refers to cascades while the dotted blue line is for tracks; for
completeness we also show the results for the full sample (solid orange
line). The numbers give the observed (above the points) and average random
value (below the points) of $N_{{\nu}}$. Figure \ref{fig:SDSS_FWHM} shows
the following:

\begin{enumerate}
\item for all samples the chance probability is strongly dependent on the
  proxy for outflow velocity.  We attribute the turn-over in p-value at
  very large velocities to small number statistics;
\item a p-value $\sim 17$ per cent is reached for ${\rm FWHM_{\rm Avg}}
  \gtrsim 800$ km s$^{-1}$ for tracks. This becomes $\sim 48$ per cent once
  the trial correction is applied;
\item a p-value $\sim 30$ per cent is reached for ${\rm FWHM_{\rm Avg}}
  \gtrsim 2,000$ km s$^{-1}$ for cascades. This becomes $\sim 60$ per cent
  once the trial correction is applied. As discussed above, we do not
  expect a signal for cascades due to the large density of SDSS sources;
\item at the ${\rm FWHM_{\rm Avg}}$ at which the p-value for tracks is
  minimum $N_{{\nu}}$ is 9, while the average value from the randomisation
  is 6.9. Even if we interpreted this excess of $\approx 2$ IceCube tracks as
  ``real'', this would imply a contribution to the IceCube signal from 
  possible AGN outflows only at the $\sim 6$ per cent level, as there 
  are 33 IceCube events in the survey area;
\item for the same ${\rm FWHM_{\rm Avg}}$ the number of SDSS sources with a
  neutrino counterpart is 26, while the whole ``parent'' SDSS sample
  includes 747 sources;
\item the p-values for the full sample are, as expected, in between those for cascades and
tracks.   
\end{enumerate}

If we split the tracks even further we get a p-value $\sim 6$ per cent for
${\rm FWHM_{\rm Avg}} \gtrsim 800$ km s$^{-1}$ for through-going
$\nu_{\mu}$, which becomes $\sim 22$ per cent with the trial correction.

\begin{figure*}
\includegraphics[height=8.0cm]{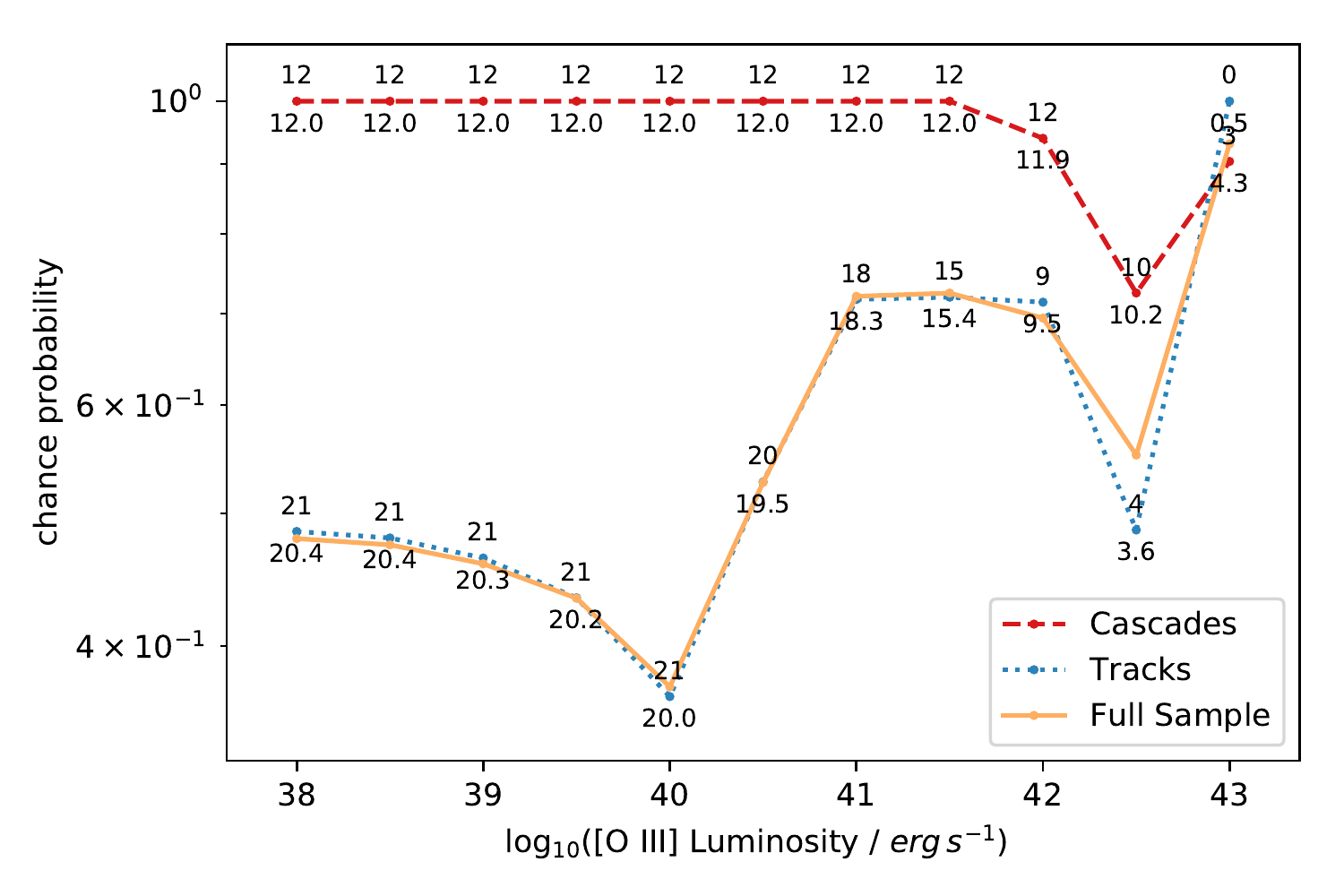}
%\includegraphics[height=8.0cm]{ALPAKA_OIII_Final.pdf}
%\vspace{0.8cm}
\caption{The chance probability of association of SDSS AGN for objects
  having $L_{[\ion{O}{iii}]}$ larger than the value on the $x$-axis with
  all IceCube events (solid orange line), cascades (dashed red line), and
  tracks (dotted blue line). The numbers give the observed (above the
  points) and average random values (below the points) of $N_{{\nu}}$.
  A p-value $\sim 37$ per cent is reached for $\log L_{[\ion{O}{iii}]}
  \gtrsim 40$ erg s$^{-1}$ for tracks.}
\label{fig:SDSS_LUM}
\end{figure*}

Figure \ref{fig:SDSS_LUM} shows the chance probability of association of
the SDSS sources with IceCube events for objects having
$L_{[\ion{O}{iii}]}$ larger than the value on the $x$-axis.  The dashed red
line refers to cascades while the dotted blue line is for tracks; the solid
orange line represents the full sample. The numbers give the observed
(above the points) and average random value (below the points) of
$N_{{\nu}}$. Figure \ref{fig:SDSS_LUM} shows the following:

\begin{enumerate}
\item for both samples, but especially for tracks, the chance probability
  depends on power;
\item a p-value $\sim 37$ per cent is reached for $\log L_{[\ion{O}{iii}]}
  \gtrsim 40$ erg s$^{-1}$ for tracks. This becomes $\sim 48$ per cent once
  the trial correction is applied;
\item a p-value $\sim 72$ per cent is reached for $\log L_{[\ion{O}{iii}]}
  \gtrsim 42.5$ erg s$^{-1}$ for cascades. This becomes $\sim 91$ per cent
  once the trial correction is applied;
\item at the $\log L_{[\ion{O}{iii}]}$ at which the p-value for tracks is
  minimum $N_{{\nu}}$ is 21, while the average value from the randomisation
  is 20.1, which means that only $\approx 1$ IceCube track might have a
  ``real'' counterpart;
\item for the same $\log L_{[\ion{O}{iii}]}$ the number of SDSS sources
  with a neutrino counterpart is 746, while the whole ``parent'' SDSS sample
  includes 22,153 sources;
\item the p-values for the full sample are, as expected, roughly in between 
those for cascades and tracks.     
\end{enumerate}

If we split the tracks even further we get a p-value $\sim 18$ per cent for 
$\log L_{[\ion{O}{iii}]} \gtrsim 41.5$ erg s$^{-1}$ for HESE tracks, which becomes 
$\sim 39$ per cent with the trial correction. 

In summary, our statistical analysis shows no significant results, although our best 
results are intriguing. We find in fact a p-value $\sim 6$ per cent for
${\rm FWHM_{\rm Avg}} \gtrsim 800$ km s$^{-1}$ for through-going
$\nu_{\mu}$ ($\sim 22$ per cent post-trial)
and $\sim 18$ per cent for $\log L_{[\ion{O}{iii}]} \gtrsim 41.5$ erg s$^{-1}$ for HESE 
tracks ($\sim 39$ per cent post-trial).

\section{Astrophysical interpretation}

Our main result is that AGN with ``bona fide'' outflows associated with an
IceCube neutrino have $\dot{\rm E}_{\rm kin}$, $\dot{\rm M}_{\rm OF}$, and
bolometric power larger than those of AGN with outflows not matched to
neutrinos (by factors $\sim 4 - 7$). The corresponding distributions are
different for the two classes of AGN at the $\gtrsim 99.6$ per cent level
for the first two parameters and at the $\gtrsim 96.2$ per cent level for
the third one. This makes perfect astrophysical sense, since AGN with
larger outflow and kinetic energy rates and bolometric powers are also
likely to be stronger neutrino emitters \citep[e.g.,][]{Wang_2016}.

One could argue that some of the presumed associations in
Tab. \ref{tab:AGN_counterparts} are not very realistic, as there are more
plausible HBL counterparts. This could certainly be the case, but it is
hard to be more quantitative about this as \cite{Padovani_2016} find a
chance probability of association with HBL only at the $\sim 1.4$ per cent
level \citep[$2.2\sigma$:][] {Resconi_2017}. A similar analysis cannot be
applied to the AGN outflow list since, although it is a very comprehensive
compilation, it does not represent a well-defined, complete sample.

We have then carried out a statistical analysis on the SDSS AGN catalogue,
finding no direct evidence of an association with the IceCube
events. However: (1) the value of the ${\rm FWHM_{\rm Avg}}$ for which we
get the smallest p-value for tracks ($\gtrsim 800$ km s$^{-1}$) is very
interesting from the astrophysical point of view, as it is above the limit
normally used to select targets for follow-up studies of AGN outflows (500
km s$^{-1}$); (2) through-going $\nu_{\mu}$ alone provide a small excess
($\sim 2\sigma$ pre-trial), which we interpret as a fluctuation but goes
in the right direction at an astrophysically relevant ${\rm FWHM_{\rm
    Avg}}$; (3) HESE tracks alone give a p-value $\sim 18$ per cent for
$\log L_{[\ion{O}{iii}]} \gtrsim 41.5$ erg s$^{-1}$ (which becomes $\sim
39$ per cent with the trial correction), which is an astrophysically
interesting value as, by selecting the upper $\sim 22$ per cent of the
$L_{[\ion{O}{iii}]}$ distribution, it points to relatively powerful AGN,
which are the most likely outflow candidates; (4) only a small fraction 
of the high ${\rm FWHM_{\rm Avg}}$ AGN in the SDSS catalogue are 
confirmed outflow sources \citep{Harrison_2014}, the majority being still 
{\it potential} outflows; (5) finally, it could also be that ${\rm FWHM_{\rm Avg}}$
is a poor proxy for outflow power, unlike $\dot{\rm E}_{\rm kin}$ and 
$\dot{\rm M}_{\rm OF}$, which depend also on the mass and the 
radius involved ($\dot{\rm E}_{\rm kin}$ =  \textonehalf $~\dot{\rm
  M}_{\rm OF}~v{^2}_{\rm max}$ and $\dot{\rm M}_{\rm OF} = 3 \times v_{\rm max}
  \times M_{\rm OF}/R_{\rm OF}$: see eq.  B.2 of \citealt{Fiore_2017}).
  That could also explain the differences between the results derived from the 
  two approaches.

Our results have two possible implications:

\begin{enumerate}

\item AGN outflows are neutrino emitters but at present we cannot get a significant 
signal from them. This could be because the neutrino and ``bona fide'' 
outflow statistics are still too low or AGN outflows are so faint that they
  cannot be revealed as point sources but contribute to the background
  neutrino emission. This would 
  explain both the results from the AGN outflow list and those from the
  SDSS catalogue. In this case AGN outflows appear to be able to explain
  only up to $\sim 6$ per cent of the IceCube signal;

\item AGN outflows are not neutrino sources. 

\end{enumerate}

Based on our results, we believe implication (ii) is not the most favoured
one at present. Further progress on this topic requires: (1) better
neutrino statistics, which will come with time as IceCube keeps taking
data. Unfortunately, the event rate is not very high (of the order of 15 yr$^{-1}$); 
(2) stacking at the positions of the outflows. This 
  can be carried out by the IceCube consortium as done, for example, for blazars 
  by \cite{ICECube17_3};
(3) a complete catalogue of AGN outflows. The main limitation of our
work, in fact, is that we either have a list of certified outflows, which
however is not a well-defined, complete sample or we have a catalogue of mostly 
{\it potential} sources. Ideally one would like to have a well-defined,
complete catalogue of ``bona fide'' AGN outflows, but this is not available
at present. 

We stress that our results do not appear to support a scenario where 
AGN outflows explain the {\it whole} IceCube signal, as suggested by 
\cite{Wang_2016} and \cite{Lamastra_2017}, but instead 
might corroborate the work of \cite{Liu_2018}, who predict a smaller
contribution. 
 
\section{Conclusions}

We have directly tested for the first time the existence of a new class of 
neutrino sources, namely matter outflows associated with AGN. We have first
cross-correlated a list of 94 ``bona fide'' AGN outflows put together by
\cite{Fiore_2017} with the only complete and updated repository of IceCube
neutrinos currently publicly available, collected by us for this purpose.
Our main result is that AGN with outflows associated with an
IceCube neutrino have outflow and kinetic energy rates and bolometric powers
larger than those of AGN with outflows not matched to
neutrinos. The corresponding distributions are also different
for the two AGN classes, significantly so for the first two parameters 
($\gtrsim 99.6$ per cent). A proper statistical analysis of this association 
cannot be carried out since the AGN outflow list, although very comprehensive, 
does not represent a well-defined, complete sample.

We have then carried out a statistical analysis on a catalogue of [O\,\textsc{iii}]~$\lambda5007$ 
line profiles using a sample of 23,264 SDSS AGN at $z < 0.4$ \citep{Mullaney_2013}, which 
can be used to determine the kinematics of the kpc-scale emitting gas. One can use the 
[O\,\textsc{iii}]~$\lambda5007$ flux-weighted average FWHM as a proxy to select AGN with 
potential outflows, together with $L_{[\ion{O}{iii}]}$. We find no significant evidence of an 
association between the SDSS AGN and the IceCube events, although the values of 
${\rm FWHM_{\rm Avg}}$ and $\log L_{[\ion{O}{iii}]}$ for which we get the smallest
p-values ($\sim 6$ and 18 per cent respectively, pre-trial) make perfect astrophysical sense. 
The former, in particular (${\rm FWHM_{\rm Avg}} \gtrsim 800$ km s$^{-1}$),  
is above the limit normally used to select targets for follow-up studies of AGN 
outflows (500 km s$^{-1}$). 

Our results are consistent with a scenario where AGN outflows are neutrino 
emitters but at present do not provide a significant signal. This
can be tested with better outflow and neutrino statistics and stacking. In any case, 
we appear to rule out a predominant role of AGN outflows in explaining the IceCube data.

\section*{Acknowledgments}

We thank Chris Harrison and Vincenzo Mainieri for useful suggestions and
discussions, Fabrizio Fiore for sending us the coordinates of the sources
in Tab. B1 of \cite{Fiore_2017}, Stefan Coenders for his contribution to
the statistical analysis and the teams, which have produced the data and
catalogues used in this paper for making this work possible. ER is
supported by a Heisenberg Professorship of the Deutsche
Forschungsgemeinschaft (DFG RE 2262/4-1). This work was supported 
by the Deutsche Forschungsgemeinschaft through grant SFB 1258 
``Neutrinos and Dark Matter in Astro- and Particle Physics''. We made use of the
TOPCAT \citep{tay05} and of the R \citep{R_2017} software packages.

%\clearpage

\label{lastpage}

% Don't change these lines
\bsp	% typesetting comment

\end{document}